# Fermi level tuning and atomic ordering induced giant anomalous Nernst effect in $Co_2MnAl_{1-x}Si_x$ Heusler alloy

Y. Sakuraba[1,2], K. Hyodo[3], A. Sakuma[3] and S. Mitani[1]

[1]Research Center for Magnetic and Spintronic Materials, National Institute for Materials Science, Sengen 1-2-1, Tsukuba, Ibaraki, 305-0047 Japan
[2] PRESTO, Japan Science and technology Agency, Saitama 332-0012, Japan
[3] Department of Applied Physics, Tohoku University, Aoba-ku, Sendai, 980-8579, Japan

**Abstract**

$Co_2MnAl$ has been predicted to have Weyl points near Fermi level which is expected to give rise to exotic transverse transport properties such as large anomalous Hall(AHE) and Nernst effects(ANE) due to large Berry curvature. In this study, the effect of Fermi level position and atomic ordering on AHE and ANE in $Co_2MnAl_{1-x}Si_x$ were studied systematically. The $Co_2MnAl$ film keeps $B$2-disordred structure regardless of annealing temperature, which results in much smaller anomalous Hall conductivity $\sigma_{xy}$ and transverse Peltier coefficient $\alpha_{xy}$ than those calculated for $L2_1$-ordered $Co_2MnAl$. Our newly performed calculation of $\sigma_{xy}$ with taking $B$2 disordering into account well reproduces experimental result, thus it was concluded that Berry curvature originating from Weyl points is largely reduced by $B$2 disordering. It was also revealed Al substitution with Si shifts the position of Fermi level and improves the $L2_1$-atomic ordering largely, leading to strong enhancement of $\alpha_{xy}$, which also agreed with our theoretical calculation. The highest thermopower of ANE of 6.1$\mu$V, which is comparable to the recent reports for $Co_2MnGa$, was observed for $Co_2MnAl_{0.63}Si_{0.37}$ because of dominant contribution of $\alpha_{xy}$. This study clearly shows the importance of both Fermi level tuning and high atomic ordering for obtaining the effect of topological feature in Co-based Heusler alloys on transverse transport properties.

Anomalous Nernst effect (ANE), which is a thermoelectric phenomenon unique to magnetic materials, has attracted attention because of several unique advantages for thermoelectric applicatioins.[1,2]   Here, the electric field of ANE ($\vec{E}_{ANE}$) can be expressed by the following equation,

$$\vec{E}_{ANE} = Q_S \left( \mu_0 \vec{M} \times \nabla T \right) , \qquad (1)$$

where $Q_S$ and $\mu_0 \vec{M}$ represent the anomalous Nernst coefficient and magnetization, respectively. As equation (1) indicates, ANE generates an electric field in the direction of the outer product of the magnetization $\mu_0 \vec{M}$ and temperature gradient $\nabla T$. This three dimensionality of ANE enables us to increase the serial voltage by using thermopiles consisting of simple laterally connected magnetic wires, because $\vec{E}_{ANE}$ appears along the surface of a heat source. This is a significant advantage for enlarging the size of the TEG module and utilizing large-area of non-flat heat sources. In addition to such attractive feature for practical applications, recent finding of large ANE originating from the materials having topological features such as $Mn_3Sn$[3] stimulated studies on ANE for gaining a fundamental understanding of the phenomenon and enhancing its thermopower.[3-16]   It has been recently reported that, ferromagnetic Heusler alloy $Co_2MnGa$ showed the largest thermopower of ANE of about 6 μV/K[6,7], which is one order of magnitude larger than that the conventional ferromagnets in Fe, Co and Ni.[17] Such large thermopower was explained as an exotic property of a magnetic Weyl semimetal in $Co_2MnGa$. Namely, large transverse thermoelectric effect intrinsically appears in $Co_2MnGa$ due to its large Berry curvature near Fermi level ($E_F$) because of the formation of Weyl points on the nodal lines of electronic bands by the spin-orbit interaction[5,7,8]. Such a topological feature of magnetic material has attracted worldwide interest for not only fundamental physics but also its great potential of practical applications. One curious issue yet to be clarified in Heusler alloy-based Weyl semimetals is how the atomic ordering and the position of $E_F$ against the Weyl points affect the sign and magnitude of anomalous Hall effect (AHE) and ANE. Since previous studies for $Co_2MnGa$ have focused on the bulk single crystal or epitaxial thin film having high $L2_1$ atomic ordering and the stoichiometric composition[6,7], it is still unclear how much ANE and AHE are sensitive to the atomic ordering and chemical composition.

  In the present study, we paid attention to $Co_2MnAl$ which is another interesting material predicted to show large intrinsic AHE[18,19] due to the existence of Weyl points near $E_F$ [20]. Previous experiment claimed the observation of large AHE in the $Co_2MnAl$ film having a random disordering of Mn and Al, so called, $B2$ disorder(Fig. 1(a),(b)).[21] However, they showed only anomalous Hall resistivity $\rho_{xy}$ as a evidence of large AHE and did not compare anomalous Hall conductivity $\sigma_{xy}$ with the theoretical value although $\sigma_{xy}$ is the intrinsic physical parameter that is theoretically accessible[22,23]. Therefore, strictly speaking, the theoretically predicted large AHE has never been confirmed in $Co_2MnAl$. It is well known that it is not easy to form $L2_1$-ordering in $Co_2MnAl$ especially in thin film[21,24]. $Co_2MnAl$ often has B2-ordering in which Mn and Al atoms are randomly disordered because of too small driving force to form $L2_1$ as indicated by very low $L2_1$ to B2 order-disorder transition temperature $T_t^{L21/B2}$ of 950K[25], which is much lower than $T_t^{L21/B2}$ in $Co_2MnGa$, 1200K. It is expected that the atomic ordering in $Co_2MnAl$ can be largely improved by the replacement of Al with Si because $L2_1$-$Co_2MnSi$ is the thermally stable intermetallic ordered compound that keeps $L2_1$ ordered structure up to its melting temperature ~ 1400K, namely, $T_t^{L21/B2}$ was estimated to be 1580K [26],that is higher than its melting point. In addition, previous study revealed that the position of $E_F$ in $Co_2MnAl$ can be tuned toward higher energy by substituting Si with Al.[27,28] Therefore, $Co_2MnAl_{1-x}Si_x$ is a suitable material to investigate how the position of Weyl points against $E_F$ and atomic ordering influences AHE and ANE.

(001)-oriented epitaxial $Co_2MnAl_{1-x}Si_x$ (CMAS) thin films having different Si:Al ratios were grown on a MgO (001) substrate using a co-sputtering technique with $Co_2MnSi$ and $Co_2MnAl$ sputtering targets. All films were deposited at ambient substrate temperature and then in-situ annealed at 600°C. The composition of the films was measured by a combination of inductively coupled plasma mass spectrometry (ICP-MS) and x-ray fluorescence analysis (XRF). In this study we made twelve CMAS thin films having a different Si:Al composition ratio $x$. The compositions of two CMAS films having nominal $x = 0$ and 0.25 were measured by ICP-MS and determined to be nearly stoichiometry $Co_{1.93}Mn_{0.98}Al_{1.08}$ and $Co_{1.88}Mn_{0.95}Al_{0.90}Si_{0.27}$ in at.%, respectively. Although we found a slight off-stoichiometry of Co and Mn compositions, we focus on the effect of Si:Al ratio $x$ on various properties in this study. Thus, the $x$ for all CMAS films is evaluated by XRF to be 0.00, 0.11, 0.14, 0.15, 0.22, 0.23, 0.29, 0.32, 0.37, 0.48, 0.51, and 1.00. For simplification, we express each CMAS films in "$Co_2MnAl_{1-x}Si_x$" using $x$ measured by XRF. The annealing temperature $T_{ann}$ dependence was studied for the films with $x = 0$ and 0.37 from 500 to 700°C to investigate the atomic ordering effect on ANE. The thickness of the films was fixed at 30 nm. The crystal structure and atomic ordering were investigated by x-ray diffraction with a Cu $K_\alpha$ source. Longitudinal and transverse electric and thermoelectric transport properties including ANE were investigated with a physical property measurement system (PPMS) for films patterned by photolithography and Ar ion milling. The electric resistivity $\rho_{xx}$ was measured using a dc four-probe method by flowing a constant dc current of 1 mA. ANE (AHE) was measured by flowing a heat (electric) current in the film plane direction and applying a magnetic field in the perpendicular direction in PPMS at 300K. As for ANE, the temperature gradient $\nabla T$ in PPMS was carefully evaluated through the following procedure : First $\nabla T$ outside of PPMS was measured using an infrared camera (InfReC R450, Nippon Avionics) for the sample with the black body coating to correct the emissivity of the samples. At the same time, the Seebeck voltage $V_{SE}$ in the film was measured outside, then the linear relationship between $V_{SE}$ and $\nabla T$ was obtained. After that, ANE voltage $V_{ANE}$ was measured together with $V_{SE}$ in PPMS, then $\nabla T$ in PPMS can be estimated through the $V_{SE}$. For a strict evaluation of Seebeck coefficient $S_{SE}$, we used the Seebeck Coefficient/Electric Resistance Measurement System (ZEM-3, ADVANCE RIKO, Inc.). We also performed a first principles calculation to evaluate $\sigma_{xy}$. The first-principles technique was the tight binding-linearized muffin-tin orbital method under the local spin-density approximation[29]. To consider the AHE effect, the spin-orbital-coupling term under the Pauli approximation was added to the non-relativistic Hamiltonian. $\sigma_{xy}$ was calculated from the Kubo-Bastin formula consisting of Fermi-surface and –sea terms[30]. Since previous theoretical studies have calculated $\sigma_{xy}$ in only ideal $L2_1$-ordered cases[18,19], in this study the electron scattering effect originating from $B2$ disorder on $\sigma_{xy}$ was taken into account in the coherent-potential-approximation[29]. About $5 \times 10^7$ $\boldsymbol{k}$-points were used for the Fermi-surface term and from $1 \times 10^6$ to $5 \times 10^7$ depending on the energy variable in the integration for the Fermi-sea term in the full Brillouin zone [30].,

Out-of-plane XRD patterns for the CMAS films annealed at 600°C are shown in Figure 1(c). We clearly detected only (002) and (004) peaks from all CMAS films, indicating (001)-oriented growth in the whole range of $x$. A clear (002) super lattice peak indicates the existence of atomic ordering between Co and (Mn,Al/Si) sites so-called $B2$ structure (Fig.1(b)). The out-of-plane lattice constant $a$, as evaluated from the (004) peak position, is plotted against $x$ in the inset of Fig. 1(c). The out-of-plane lattice constant for $Co_2MnAl$ and $Co_2MnSi$ films are 5.74 and 5.63 Å, respectively, which is similar with the reported values in literature, 5.755 and 5.654 Å.[26] The $a$ almost linearly decreases with increasing Si composition ratio, following Vegard's law, indicating a formation of single phase CMAS in whole range of $x$. We also measured the (111) super lattice peak arising $L2_1$-ordered structure by tilting the film plane to 54.7° from the normal direction. No (111) peak appears from $x = 0$ to 0.15, but tiny detectable peak is observed from $x = 0.22$ to 0.51 as shown in Figure1(d). The (111) peak intensity appears to be larger with increasing Si composition ratio and the strongest peak was observed in $Co_2MnSi$, which can be explained by enlargement of $T_t^{L21/B2}$ by the substitution of Si with Al in $Co_2MnAl$ as mentioned earlier. Recent study clearly found that the sign of AMR in Co-based Heusler is sensitive to the position of $E_F$ inside/outside the energy gap in minority spin channel(half-metalic gap), namely, the sign of AMR is negative(positive) when $E_F$ is inside(outside) of half-metallic gap.[28,31,32] Therefore we measured AMR for our CMAS films to see the change of $E_F$ position indirectly and found the clear sign change from positive to negative from CMA to CMS at around $x = 0.4$.(see Supplementary information). This result supports the $E_F$ shifting toward higher energy by replacing Si with Al as we expected.

Figure 2(a) shows the perpendicular magnetic field dependence of the anomalous Hall resistivity $\rho_{yx}$ for the CMAS thin films measured at 300 K. The $Co_2MnAl$ film had the largest $\rho_{yx}$ of about +18 μΩ·cm which is very close value with the previous study[21]. $\rho_{yx}$ almost monotonically decreases upon replacing Al with Si, as shown in Figure 2(b), and the $Co_2MnSi$ film had the smallest $\rho_{yx}$, 0.7 μΩ·cm. This result well agrees with the previous study of Hall effect in $Co_2MnSi_{1-x}Al_x$ polycrystalline bulk samples reported by Prestigiacomo et al.[33] The longitudinal conductivity $\rho_{xx}$ shown in Fig.2(b) is nearly constant of about 240-260 μΩcm in the region of $x = 0$ to 0.32, and then reduces down to 83 μΩcm from 0.37 to 1.00, which must be more or less related with the improvement of $L2_1$-ordering with Si. The anomalous Hall angle $\theta_{AHE}$ and anomalous Hall conductivity $\sigma_{xy}$ are evaluated using the equations $\theta_{AHE} = \rho_{yx}/\rho_{xx}$ and $\sigma_{xy} = \rho_{yx}/(\rho_{xx}^2 + \rho_{yx}^2)$, respectively, and plotted in Figs.2(d) and (e). $\theta_{AHE}$ clearly monotonically decreases with increasing Si; $Co_2MnAl$ showed the largest magnitude of anomalous Hall angle, $|\theta_{AHE}|$ of 7.3%, whereas $|\theta_{AHE}|$ decreases with $x$ to 0.8% in Co$_2$MnSi. It should be noted here that the $\sigma_{xy}$ obtained for Co$_2$MnAl and Co$_2$MnSi are 295(362) and 96(101) S/cm at 300K(10K), respectively, which are lower than the calculated intrinsic contribution of AHE, $\sigma_{xy}^{int}$, 1265 and 193 S/cm for $L2_1$-ordered Co$_2$MnAl and Co$_2$MnSi[19]. Since the theoretical intrinsic mechanism contribution for AHE does not take any electron scattering effect into consideration, experimentally observed $\sigma_{xy}$ in thin films is reduced even at low temperature by unavoidable scatterings at the surface/interface such as the film surface, film/substrate interface, and grain boundaries. As the $\sigma_{xy}$ in Fe epitaxial film reduces with decreasing its thickness[34], the existence of electron scattering can be one reason for $\sigma_{xy} < \sigma_{xy}^{int}$. However, the deviation between $\sigma_{xy}$ and $\sigma_{xy}^{int}$ for Co$_2$MnAl seems too large (see Figure 2(e)) to be explained by such an additional scattering. To understand this mechanism, we calculated density of state (DOS) and $\sigma_{xy}^{int}$ for no only $L2_1$ and but also $B2$ Co$_2$MnAl. As shown in Fig.3(c), calculated $\sigma_{xy}^{int}$ in $L2_1$-Co$_2$MnAl exhibits large variation from 300 to 1600S/cm within even small $\pm 0.3$eV range around $E_F$ and takes large value of 931S/cm at $E_F$. On the other hand, disordered $B2$-Co$_2$MnAl was predicted to show much smaller $\sigma_{xy}^{int}$, 258S/cm, at $E_F$ with small slope against energy. As shown in Fig.2(e), this $\sigma_{xy}^{int}$ for B2-Co$_2$MnAl is close to the experimental $\sigma_{xy}$. Although it has not been elucidated by our calculation that how Berry curvature in the momentum space changes from $L2_1$ to $B2$ disordering structures, it is expected the $B2$ disorder smears the whole band dispersion including the bands forming the Weyl points, which must reduce the Berry curvature near Fermi level. This smearing effect of band dispersion can be seen from the blurred total DOS of B2 structure compared to the sharp DOS of $L2_1$ as shown in Figs.3(a) and 3(c). Thus, it is concluded that observed small $\sigma_{xy}$ in B2-Co$_2$MnAl film is attributed to this intrinsic reduction of $\sigma_{xy}^{int}$ from $L2_1$ to B2. To see Si substitution effect, we also calculated DOS and $\sigma_{xy}^{int}$ for $L2_1$ and $B2$ Co$_2$MnAl$_{0.63}$Si$_{0.37}$ as shown in Figs.3(b) and (d), respectively. If one compares the DOS in $L2_1$-ordered Co$_2$MnAl and Co$_2$MnAl$_{0.67}$Si$_{0.33}$ shown in Figs.3(a) and (b), it is clearly confirmed that the $E_F$ shifts by about +0.2 eV with keeping the shape of DOS near $E_F$. Because of this shift of $E_F$, the peak of $\sigma_{xy}^{int}$ we can see near $E_F$ in $L2_1$-Co$_2$MnAl appears at around -0.22 eV in Co$_2$MnAl$_{0.63}$Si$_{0.37}$. Consequently, we can see small difference of $\sigma_{xy}^{int}$ between $L2_1$ and $B2$, 370 and 268 S/cm, in Co$_2$MnAl$_{0.63}$Si$_{0.37}$, which can be an explanation for observed small $\sigma_{xy}$ in the Co$_2$MnAl$_{0.63}$Si$_{0.37}$ film and other CMAS films of $x$ = 0.22-0.51 regardless of their partial $L2_1$-ordering.

The $x$ dependence of thermopower of ANE and Seebeck effect ($S_{ANE}$ and $S_{SE}$, respectively) are summarized in Figure 4(a) and (b). Interestingly, the Co$_2$MnAl film that showed the largest AHE exhibits a small $S_{ANE}$ of +0.9 μV/K, and $S_{ANE}$ gradually grows as more Al is substituted with Si. The largest $S_{ANE}$ of +3.9 μV/K was observed for Co$_2$MnAl$_{0.63}$Si$_{0.37}$. Above $x$ = 0.37 the $S_{ANE}$ reduces with following $x$, finally drops down to + 0.7 μV/K for Co$_2$MnSi. On the other hand, the sign of $S_{SE}$ is negative in the whole range of $x$. With increasing $x$, the magnitude of $S_{SE}$ gradually increases with $x$ from -7.7 μV/K in Co$_2$MnAl to -21.1 μV/K in Co$_2$MnAl$_{0.49}$Si$_{0.51}$, then decreases to -11.7 μV/K in Co$_2$MnSi. Therefore, we found that the $x$ for the highest AHE, Seebeck effect and ANE are different in prepared CMAS films. Here we analyze the ANE in CMAS using a following linear response equation of $S_{ANE}$,

$$S_{ANE} = \rho_{xx}\alpha_{xy} + \rho_{xy}\alpha_{xx} \qquad (2)$$

Here $\alpha_{xx}$ and $\alpha_{xy}$ are the longtudinal and transverse Peltier coefficient, respectively. Eq.(2) tells us that there are two different phenomenal sources in ANE. For simplifying the following explanation, we denote the first and second terms as $S_I = \rho_{xx}\alpha_{xy}$ and $S_{II} = \rho_{xy}\alpha_{xx}$, respectively. Since $S_{SE} = \rho_{xx}\alpha_{xx}$, $S_{II}$ can be converted to $S_{SE} \cdot \theta_{AHE}$, therefore, $S_{II}$ is regarded as the contribution of AHE on ANE induced by a Seebeck-driven longitudinal current. On the other hand, $S_I$ originates from the direct conversion from the temperature gradient to transverse current via $\alpha_{xy}$ as expressed in $\alpha_{xy}\nabla T = i_{xy}$. Figure 4(c) plots $S_{II}$ estimated from observed $S_{SE}$ and $\theta_{AHE}$ against $x$. Although we observed a large difference of $x$ dependence between the magnitudes of AHE and ANE in the CMAS films, the trend of $x$ dependence of $S_{II}$ is similar to that of $S_{ANE}$. An important point here is that the magnitude of $S_{II}$ is smaller than the observed $S_{ANE}$ in whole range of $x$. Thus the remaining part of $S_{ANE}$ would arise from $S_I$ by following the eq.(2). Evaluated $S_I$ is plotted in Figure 4(c). It can be seen that the contribution of $S_I$ is larger than $S_{II}$ except for Co$_2$MnAl. Particularly, the largest $S_{ANE}$ of +3.9 μV/K at $x$ = 0.37 arises from the constructive but dominant contribution of $S_I$ (+3.0 μV/K) against $S_{II}$ (+0.9 μV/K). $\alpha_{xy}$ evaluated from $\alpha_{xy} = S_I/\rho_{xx}$ is plotted in Fig.4(d). It is clearly appeared that $\alpha_{xy}$ becomes larger by replacing more Al with Si, indicating that $\alpha_{xy}$ is sensitive to the position of Fermi level and atomic ordering. The $\alpha_{xy}$ for Co$_2$MnAl and Co$_2$MnAl$_{0.63}$Si$_{0.37}$ films are 0.14 and 1.18A/mK, respectively. $\alpha_{xy}$ originating from intrinsic contribution of AHE, $\alpha_{xy}^{int}$, can be theoretically evaluated from the energy dependence of $\sigma_{xy}^{int}$ using the following Mott's relation based on classic Boltzmann equation. [35]

$$\alpha_{xy}^{int} = \frac{1}{-eT}\int_{-\infty}^{\infty} \sigma_{xy}^{int}(\varepsilon)(\varepsilon - E_F)\left(-\frac{df}{d\varepsilon}\right)d\varepsilon \qquad (3)$$

To obtain $\alpha_{xy}^{int}$ at 300K, we calculated $\alpha_{xy}^{int}$ by setting $\pm 260 meV (= 0.02Ry)$ as the integration range of this calculation which is enough large for the term of $\frac{df}{d\varepsilon}(\varepsilon, T)$ to have a finite value at 300K. As shown in Fig. 4(d), calculated $\alpha_{xy}^{int}$ for $L2_1$ and $B2$-Co$_2$MnAl(Co$_2$MnAl$_{0.63}$Si$_{0.37}$) are 0.92(3.08) and 0.47(0.57), respectively. Simply speaking, $\alpha_{xy}^{int}$ is sensitive to the shape and slope of $\sigma_{xy}^{int}$ at around $E_F$, namely, even(odd) function-like behavior leads to small(large) $\alpha_{xy}^{int}$. As we can see in Fig.3(c), $L2_1$-Co$_2$MnAl shows nearly even function like behavior around $E_F$ within the integration range, whereas, $L2_1$- Co$_2$MnAl$_{0.63}$Si$_{0.37}$ shows odd function like behavior with a large negative slope, which is a reason for much larger $\alpha_{xy}^{int}$ in $L2_1$-Co$_2$MnAl$_{0.63}$Si$_{0.37}$. In B2 disordered case, because $\sigma_{xy}^{int}$ shows very small change in both Co$_2$MnAl and Co$_2$MnAl$_{0.63}$Si$_{0.37}$, $\alpha_{xy}^{int}$ was estimated to be very small. As can be seen in Fig.4(d), experimental $\alpha_{xy}$ for Co$_2$MnAl$_{0.63}$Si$_{0.37}$ is close to the calculated value for $B2$-case but reasonably located in between $L2_1$ and $B2$. Therefore, we concluded that the enlargement of $\alpha_{xy}$ in our CMAS thin films is attributed to not only Fermi level shifting but also the improvement of $L2_1$-atomic ordering by Si substitution for Al.

To see the effect of atomic ordering more clearly, we investigated the annealing temperature $T_{ann}$ dependence of atomic ordering, AHE and ANE in the Co$_2$MnAl and Co$_2$MnAl$_{0.63}$Si$_{0.37}$ films. Figure 5(a) shows the intensity ratio of (111) super lattice peak to (004) fundamental peak, $I_{111}/I_{004}$, against $T_{ann}$. Co$_2$MnAl film does not show (111) peak even after annealing at 700°C, indicating that Co$_2$MnAl keeps B2 disordered structure regardless of $T_{ann}$. Therefore, $\sigma_{xy}$ in Co$_2$MnAl film is around 300S/cm and shows no remarkable variation against $T_{ann}$ (Fig.5(b)). In contrast, tiny (111) that appears at $T_{ann}$ = 600°C in Co$_2$MnAl$_{0.63}$Si$_{0.37}$ was strongly enlarged by increasing $T_{ann}$, up to 650°C.

Oppositely, (111) peak does not appears at 500 °C . The order parameter $S_{L21}$ of the $L2_1$ structure was evaluated as 0.48 at 650°C from $S_{L21}^2 = (I_{111}^{obs}/I_{004}^{obs})/(I_{111}^{sim}/I_{004}^{sim})$, where $I_{hkl}^{obs/sim}$ is the observed/simulated integrated diffraction intensity of the (hkl) peak, indicating that there is nearly 50% mixture phases of *B*2 and *L*2$_1$ at 650°C. Observed $\sigma_{xy}$ in Co$_2$MnAl$_{0.63}$Si$_{0.37}$ gradually increases with $T_{\text{ann}}$ (Fig.5(b)) from 136 S/cm at 500°C to 275 S/cm at 700°C , whose tendency is in qualitative agreement with the calculated $\sigma_{xy}^{int}$ shown in Fig.3(d). A drastic increase of $S_{\text{SE}}$ and $S_{\text{ANE}}$ were also observed in the Co$_2$MnAl$_{0.63}$Si$_{0.37}$ annealed at above 650°C(Fig.5(c)-(e)) in contrast to no remarkable change of them in the Co$_2$MnAl aginst $T_{\text{ann}}$, indicating that the enlargement of both $S_{\text{SE}}$ and $S_{\text{ANE}}$ arises from the improvement of *L*2$_1$-ordering. The highest $S_{\text{ANE}}$[6,7] of 6.1μV/K, which is comparable to the previous reports in Co$_2$MnGa[6,7], was observed for the Co$_2$MnAl$_{0.63}$Si$_{0.37}$ film annealed at 650°C. At the same time, as can be seen in Fig.5(f), $\alpha_{xy}$ reaches 1.8A/mK at 650°C which is reasonably in between theoretical $\alpha_{xy}^{int}$ for *L*2$_1$ and *B*2. Such large $\alpha_{xy}$ gives rise to dominant $S_{\text{I}}$ contribution of 4.4 μV/K for the total ANE of 6.1μV/K. Therefore, it is suggested that giant ANE in CMAS film achieved in this study is due to both the Fermi level shifting and improvement of *L*2$_1$-atomic ordering in Co$_2$MnAl which has been predicted as a Weyl semi-metal.

In conclusion, in this work, we studied anomalous Hall and Nernst effect in the Co$_2$MnAl$_{1-x}$Si$_x$ from both experiment and first-principle calculation to see the effect of Fermi level position and the degree of atomic ordering on AHE and ANE. It was clearly confirmed that Fermi level shifts toward higher energy and *L*2$_1$-ordering improves with increasing Si composition ratio *x*. Observed $\sigma_{xy}$ in the Co$_2$MnAl film having no *L*2$_1$-ordering is much smaller than the calculated intrinsic $\sigma_{xy}$ for *L*2$_1$-orderd structure but close to our calculation for B2-disordered structure, suggesting that theoretically predicted large AHE due to the existence of Weyl points in Co$_2$MnAl is weaken by unavoidable B2 disordering in reality. Although Al substitution with Si does not strongly affect $\sigma_{xy}$, the transverse Peltier coefficient $\alpha_{xy}$ was clearly enlarged with increasing Si, and the highest $\alpha_{xy}$ was obtained in Co$_2$MnAl$_{0.63}$Si$_{0.37}$. As predicted by our calculation, $\alpha_{xy}$ was enlarged by improving *L*2$_1$-ordering in Co$_2$MnAl$_{0.63}$Si$_{0.37}$, and finally giant thermopower of ANE of 6.1μV/K was achieved in Co$_2$MnAl$_{0.63}$Si$_{0.37}$ film having the mixture of *L*2$_1$- and *B*2-phase. Our result indicates that, both Fermi level tuning and high atomic ordering is critically important to realize exotic transverse transports in Co-based Heusler Weyl semi-metals. This knowledge will be beneficial for a future material development to realize practical thermoelectric applications using ANE.

**Acknowledgements**

Authors thank K. Masuda, K. Sumida, A. Kimura, K. Takanashi, S. Maekawa, K. Hono, and K. Uchida for valuable discussions and N. Kojima and H. Ikeda for a technical support. This work was supported by a JSPS KAKENHI Grant-in-Aid for Young

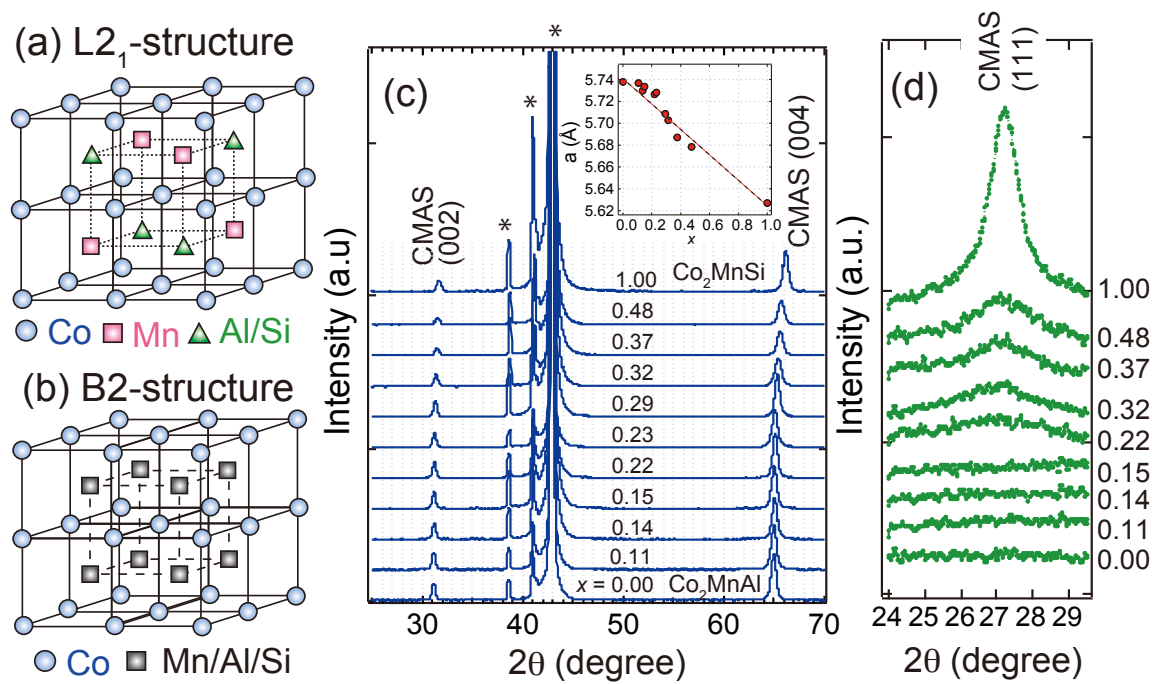

**Figure 1** (a),(b) The schematic views of $L2_1$- and $B2$ structure of CMAS. (c) Out-of-plane XRD patterns for $Co_2MnAl_{1-x}Si_x$ thin films annealed at 600°C. The peaks denoted by * originate from MgO (001) substrate. The inset shows Si composition $x$ dependence of lattice constant $a$ evaluated from position of (004) peak position. (d)XRD patterns measured by tilting the film normal plane by 54.7°to see (111) superlattice peak.

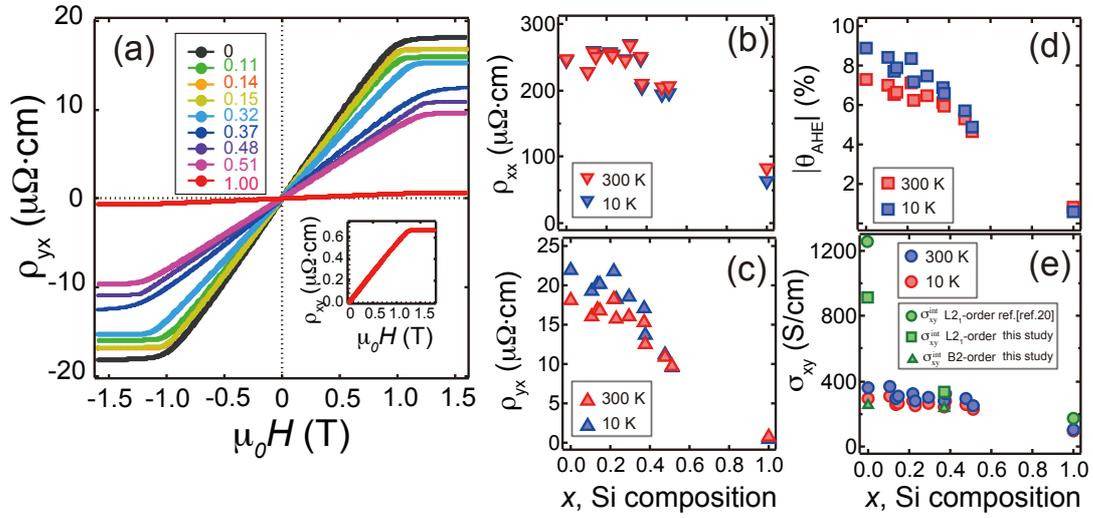

**Figure 2** (a) Perpendicular magnetic field $H$ dependence of anomalous Hall resistivity $\rho_{xy}$ for $Co_2MnAl_{1-x}Si_x$ thin films measured at 300 K. Si composition dependence of $\rho_{xx}$ (b), $\rho_{yx}$ (c), anomalous Hall angle $|\theta_{AHE}|$ (c) and $\sigma_{xy}$ (c). The data measured at 10 and 300K are shown in (b)-(e). Theoretical $\sigma_{xy}^{int}$ for $L2_1$-and $B2$- $Co_2MnAl$ and $Co_2MnAl_{0.67}Si_{0.33}$ are also plotted in (e).

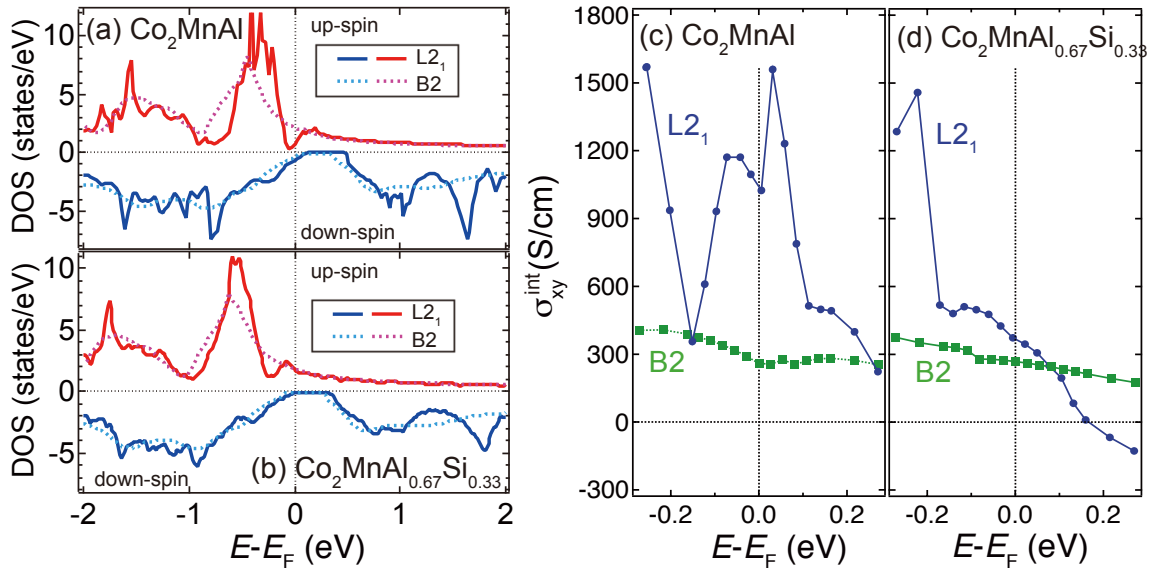

**Figure 3** First principles calculation of the spin-resolved DOS for $Co_2MnAl$(a) and $Co_2MnAl_{0.67}Si_{0.33}$(b). For both compositions we calculated the DOS in $L2_1$ and $B2$ disordered structure. (b) Calculated energy dependence of $\sigma_{xy}$ for B2 and $L2_1$-ordered $Co_2MnAl$(c) and $Co_2MnAl_{0.63}Si_{0.37}$(d).

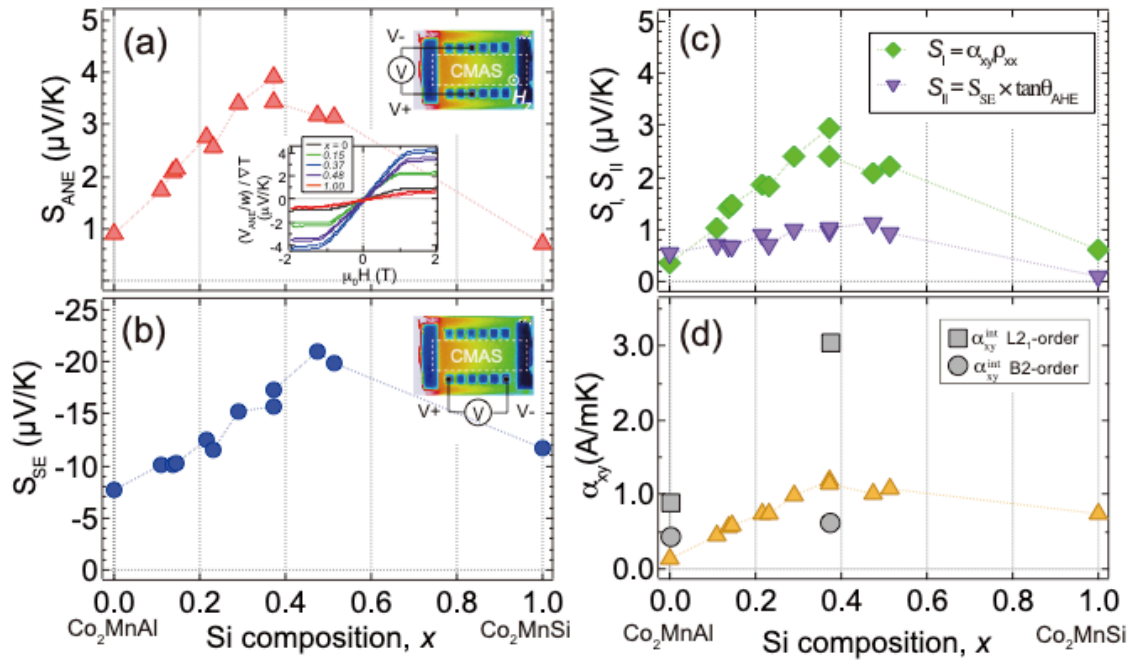

**Figure 4** Si composition $x$ dependence of $S_{ANE}$ (a), $S_{SE}$ (b), $S_I$ and $S_I$ (c) and $\alpha_{xy}$(d). The inset of (b) shows the external magnetic field dependence of $V_{ANE}$ normalized by the sample width $w$ and given temperature gradient $\nabla T$ in CMAS films. Theoretically calculated $\alpha_{xy}^{int}$ using eq.(3) are also plotted in (d)

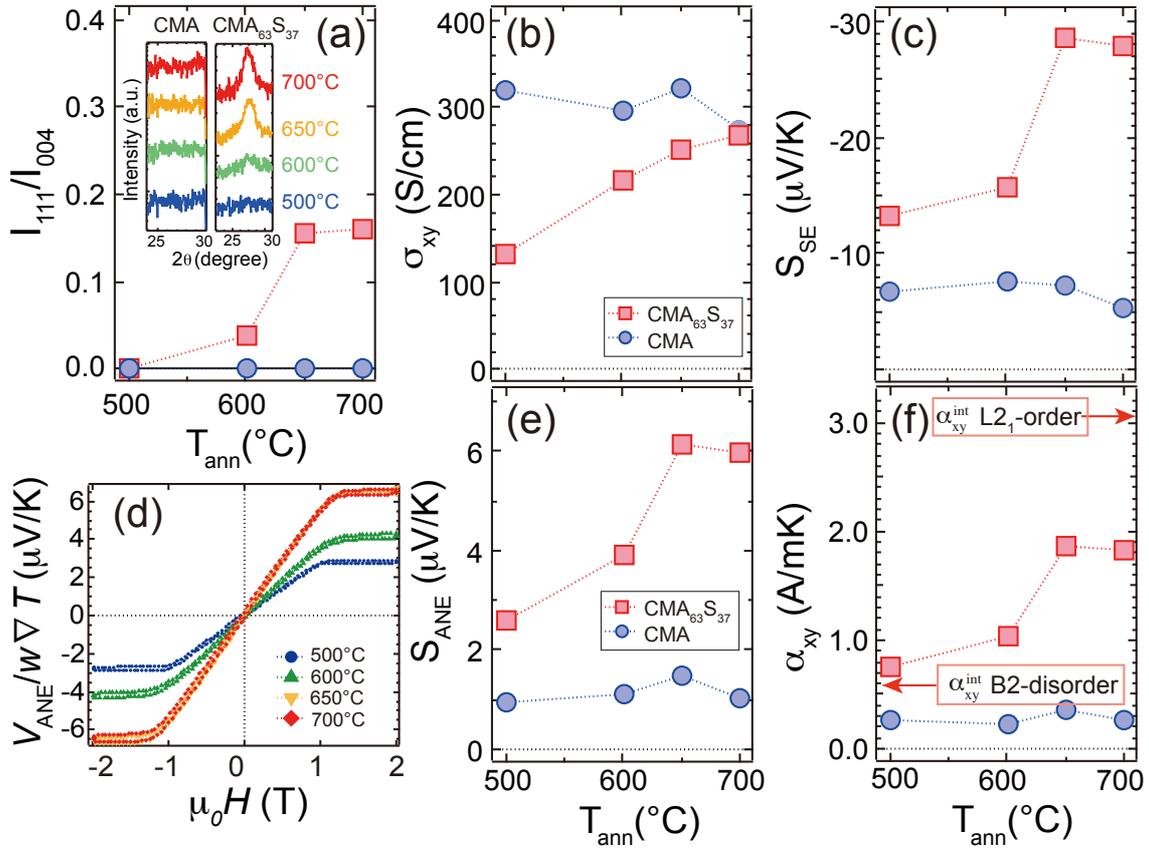

**Figure 5** (a) Intensity ratio of XRD diffraction (111) to (004) peaks against $T_{ann}$. Inset shows XRD patterns in Co$_2$MnAl and Co$_2$MnAl$_{0.63}$Si$_{0.37}$ thin films for the 2θ range of $L2_1$-super lattice (111) peak. $T_{ann}$ dependence of $\sigma_{xy}$, $S_{SE}$, $S_{ANE}$ and $\alpha_{xy}$ in Co$_2$MnAl and Co$_2$MnAl$_{0.63}$Si$_{0.37}$ are shown in (b),(c), (e) and (f), respectively. The arrows in (f) is the theoretical $\alpha_{xy}^{int}$ for $L2_1$ and $B2$ Co$_2$MnAl$_{0.63}$Si$_{0.37}$. (d) External magnetic field dependence of $V_{ANE}$ normalized by the sample width $w$ and given temperature gradient $\nabla T$ in Co$_2$MnAl$_{0.63}$Si$_{0.37}$ annealed at different temperature.